**Assessing Gaze and Pointing: Human Cue Interpretation by Indian Free-Ranging Dogs in a Food Retrieval Task**

Srijaya Nandi[a], Dipanjan Roy[a], Aesha Lahiri[a], Anamitra Roy[a], Anindita Bhadra[a*]

[a]Department of Biological Sciences, Indian Institute of Science Education and Research, Kolkata, Mohanpur, Nadia, West Bengal, India

Corresponding author

Prof. Anindita Bhadra

abhadra@gmail.com

**Abstract**

The urban habitat provides a landscape that increases the chances of human-animal interactions, which can lead to increased human-animal conflict, but also coexistence. Some species show high levels of socio-cognitive abilities that enable them to perceive communicational gestures of humans and use them for their own benefit. This study investigated the ability of Indian free-ranging dogs (*Canis lupus familiaris*) to utilise human social-referential cues (pointing and gazing) to locate hidden food, focusing on the relative effectiveness of unimodal versus multimodal cues. A total of 352 adult free-ranging dogs were tested in an object-choice task involving six different cue conditions: control (no cue), negative control (one baited bowl, no cue), combined pointing and gazing, pointing-only, gazing-only, and conflicting cues (pointing and gazing at opposite bowls). The dogs successfully chose the correct target only in the combined pointing and gazing condition, while performance under unimodal and conflicting cue conditions did not differ significantly from chance. This highlights the importance of signal redundancy and clarity in interspecific communication for this population. A dog's demeanor was a significant predictor of its willingness to engage: affiliative dogs were significantly more likely to succeed in the overall experiment and displayed a significantly shorter approach latency compared to anxious and neutral dogs. While demeanor affected the approach latency, it did not affect the accuracy of the choice, decoupling the dogs' personality from its cognitive ability to comprehend the clear cue. Neither the dogs' sex nor the experimental condition significantly predicted approach latency. These findings reaffirm that Indian free-ranging dogs are adept at interpreting human social signals, and reveals that their success is contingent upon the multimodal clarity of the signal. This underscores the complex interplay between signal characteristics, individual temperament, and cognitive performance in dog-human communication.

# Introduction

Communication is a fundamental capability that enables animals to pass information between individuals, facilitating complex social interactions and activities essential for survival (Seyfarth et al., 2010). At a minimum, communication requires two participants: a sender and a receiver. As defined by C. Cherry (1966), communication is "the establishment of a social unit from individuals by the use of language or signs", underscoring its role in the emergence of social life among animal species. While communication is not limited to social or group-living species, it is an essential pre-requisite for the formation and maintenance of social groups, and can thus be considered as one of the pillars of social evolution. It should be noted that communication, from simple calls to complex songs and languages, primarily serves the purpose of intra-specific exchange of information (Xie et al., 2024; Garcia and Favaro, 2017).

Among the various forms of animal communication, referential gestures which are purposeful body movements intended to evoke specific goal-oriented behaviours in the recipient, are particularly significant, yet rare, occurring most frequently in species with complex social systems such as those characterized by fission-fusion dynamics (Hobaiter & Byrne, 2011; Emery, 2000; Pika, 2012). Pointing and gaze following are communicative cues closely associated with reference, a foundation of human language (Lyons, 1975). The ability to use referential cues, such as gazing and pointing, is a cornerstone of complex social communication. Other species are not known to possess these complex communicative skills. However, animals that have a close association with humans, as domestic animals, are attributed the skills of understanding human attentional states, gestural communication and even spoken languages (Ringhofer et al., 2021; Langner et al., 2023; Schünemann et al., 2021). One historic case is of Clever Hans, a horse thought to possess extraordinary cognitive skills. It was later revealed that these were rooted in the animal's sensitivity to subtle human cues (Pfungst, 1911). The ability of non-human animals to perceive and comprehend human communicative cues has been a matter of curiosity as well as debate, and deserves the attention it receives from the scientific community.

Research across multiple species reinforces that communication abilities are shaped by the interplay between social environment and evolutionary history (Garcia et al., 2020; Whiten and Schaik, 2007). The domestic dog, *Canis lupus familiaris*, shares a very unique relationship with our species. Humans and dogs have co-evolved for generations, and dogs have perhaps undergone a process of self-domestication to earn the title of "Man's best friend" (Ostrander and Giniger, 1997). Due to this shared history, dogs offer an ideal model system for exploring the abilities of animals to follow human communicative signals (Paxton, 2000; Schleidt, 1998). They not only perceive human referential gestures but also use such gestures themselves (Worsley and O'Hara, 2018). Experimental evidence supports the domestication hypothesis, positing that domestication drives dogs' heightened social-communicative skills compared to their wild relatives, such as wolves (Hare et al., 2002). Miklósi et al. (2003) have found that dogs can follow human gaze while wolves fail to do so, since they never look at human faces even when faced with an unsolvable task. There is also evidence of the enculturation hypothesis, which states that human influence has a huge role to play on the ability of animals to follow human directional cues (Call and Tomasello, 1996). Numerous studies, for instance, have demonstrated that pet dogs can follow human gaze to locate desired food or objects (Kaminski et al., 2009, 2013; Miklósi et al., 1998).

Despite this general consensus, the relative effectiveness of different human cues, and the robustness of gaze-following itself, remains a topic of investigation. A central point of discussion is the comparison between gaze and pointing. Several studies have reported that pet dogs generally perform better and respond more effectively when following pointing cues than when following gaze cues alone (Ittyerah & Gaunet, 2008; McKinley & Sambrook, 2000; Miklósi et al., 1998). Furthermore, some research directly challenges the functional success of gaze-following. Duranton et al. (2017) showed that while pet dogs can physically follow human gaze, their choices in related object-selection tasks did not exceed chance levels. This contrasts with other findings suggesting dogs are capable of performing above chance, particularly when the cue involves both head orientation and a direct gaze at the target (Soproni et al., 2001).

The complexity of this socio-cognitive skill is highlighted by its variability, both within and between species. The specific presentation of the cue, for example, heavily influences a dog's success. Pet dogs are more likely to follow a head turn if it is preceded by ostensive communicative signals, such as eye contact, demonstrating a functionally infant-like social competence (Téglás et al., 2012).

Unlike pets, free-ranging dogs have no owners; their movements and activities are self-directed (Boitani and Ciucci, 1995), yet they rely heavily on resources generated from human habitations (Sen Majumder et al., 2014). They constitute roughly 80% of the global dog population (Hughes and Macdonald, 2013) and thrive on scavenging, begging, and handouts from humans (Bhadra and Bhadra, 2014; Sen Majumder et al., 2014). Humans often use pointing, head orientation and eye gaze to indicate the direction of their attention when interacting with free-ranging dogs on the streets. While previous research has assessed free-ranging dogs' ability to follow pointing gestures paired with gazing in finding hidden food (Bhattacharjee et al., 2020), the role of pointing and gazing as independent cues in object choice tests remains understudied.

This study aims to address this gap by investigating the relative roles of human pointing and gazing as communicative cues in enabling free-ranging dogs to locate hidden food. It seeks to examine whether gaze and pointing alone, or in combination, influence the behaviour of free-ranging dogs in their everyday interactions with humans. We also seek to understand the capability of free-ranging dogs to find hidden food when these cues are directed in conflicting directions. We hypothesize that pointing would be more impactful as a means of communication with free-ranging dogs, than gaze.

**Methodology**

**Ethical Approval**

The study design did not violate the Animal Ethics regulations of the Government of India (Prevention of Cruelty to Animals Act 1960, Amendment 1982). The experimental protocol was approved by the IISER Kolkata Animal Ethics Committee, as part of a larger project sanctioned by the SERB (EMR/2016/000595).

**Subjects**

A total of 352 adult free-ranging dogs were randomly selected in various urban and semi-urban locations in West Bengal, India. Adult status for males was confirmed by observing descended testes, while for females, it was determined by the presence of darker-shaded nipples; gender was identified through observation of the genitalia. Each selected location was visited only once, ensuring that each dog participated in the study a single time. As the dogs were free-ranging, their hunger status could not be established or controlled. However, we avoided conducting experiments during the late afternoon, when humans generally feed dogs in West Bengal (findings from ongoing surveys).

**Study Areas**

Experiments were conducted within the cities and towns of Kalyani, Kanchrapara, Gayeshpur, Kataganj, Jaguli, Anandanagar, Balindi, Barrackpore, Chakdaha, Naihati, Ranaghat and Shyamnagar in West Bengal, encompassing a representative mix of urban and semi-urban habitats (Figure S1).

**Experimenters**

The research involved two experimenters: Experimenter 1 (E1), a male, and Experimenter 2 (E2), a female. During the familiarization phase, E1 administered the procedure while E2 recorded the session. In the test phase, E2 delivered the experimental cues while E1 handled the recording.

**Experimental Setup**

The apparatus used throughout comprised of blue plastic bowls, each with a volume of approximately 500 ml. The familiarization phase utilized a single bowl, while the test phase involved two identical bowls placed approximately one meter apart. Cardboard pieces were used to cover the bowls, in order to hide the food placed inside the bowls.

**Experimental Procedure**

**Familiarization Phase**

During familiarization, solitary dogs encountered on the street were presented with a piece of raw chicken weighing about ten grams, brought close to the dog's nose (within a maximum distance of 5 cm) for two to three seconds. E1 then exhibited the chicken to the dog, placed it inside the plastic bowl, and covered it with a piece of cardboard, maintaining eye contact with the dog during baiting to secure its attention. The bowl was placed 1–1.5 meters from the dog, and E1 withdrew to a position approximately 0.5 meters away, while E2 filmed the proceedings. Success in this phase was marked by the dog's ability to remove the cardboard and consume the chicken. Dogs failing to interact were given up to three familiarization trials. Only those dogs that succeeded in eating the chicken, proceeded to the test phase. Before placing the bowl, ostensive cues were provided to the dogs by means of eye contact and then calling using a commonly used positive vocalization "ae-ae-ae" (Bhattacharjee et al., 2017).

**Test Phase**

For the test phase, E2 arranged two bowls approximately one meter apart and about two meters from the dog, then stood approximately 0.5 meters from the midpoint separating the bowls. The distances maintained were the same as those used by Soproni et al. (2001). E1 moved aside, allowing E2 to assume one of five randomly assigned postures:

**a) Condition 1:** no pointing or gazing (control)

**b) Condition 2:** control with only one bowl containing chicken (negative control)

**c) Condition 3:** both pointing and gazing at the same bowl

**d) Condition 4:** pointing-only at one bowl

**e) Condition 5:** gazing-only at one bowl

**f) Condition 6:** pointing at one bowl and gazing at the other.

Eye contact was established with the dog, and E2 then assumed the predetermined randomly assigned posture and called the dog three times using the same positive vocalization "ae-ae-ae". The experimenter maintained the cueing posture for a maximum of 60 seconds. If the dog did not approach a bowl within this time, the trial was terminated. A maximum of three experiments was conducted for a particular dog, if the dog failed to approach during this phase. If the dog did approach a bowl, the posture was held until the dog had consumed the chicken pieces from both the bowls, or for a maximum of 180 seconds following the first approach (to the first bowl chosen), whichever occurred first. The kind of pointing cue used for this experiment was a dynamic distal cue, similar to the one used by Bhattacharjee et al. (2020).

For postures involving cues, the bowl side for gazing or pointing was counterbalanced across trials. The control posture involved standing neutrally, facing the horizon with hands at the sides. The negative control was similar to the control, with the exception that only one of the bowls contained chicken. Each dog was tested only once to ensure the independence of data. The control condition was performed to test for any side preference in dogs. The negative control condition was done to see if the choice of the dogs was affected by odour.

**Behavioural Coding and Parameters**

All experimental trials were video-recorded for subsequent analysis. A single, trained observer (SN) coded all videos to extract the following behavioural parameters:

**Side of first approach:** For the control (Condition 1) and the incongruent cueing condition (Condition 6), the first bowl (left or right) the dog approached was recorded.

**Choice (Correct/Incorrect):** For the negative control (Condition 2), congruent cueing (Condition 3), pointing-only (Condition 4) and gazing-only (Condition 5), the dog's choice was coded as correct (1) or incorrect (0). A choice was defined as "correct" if the dog approached the bowl indicated by the referential cue, or in the negative control condition, the bowl that contained the chicken piece.

**Approach latency:** This was measured as the time in seconds from the experimenter's first call until the dog approached a bowl. Latency was calculated for both the familiarization and test phases. During the test phase, latency to the first and second bowls was recorded. Latency to the second bowl was calculated as the time of approach to the second bowl minus the time of approach to the first. Approach latency was recorded in milliseconds during video coding, providing the necessary temporal resolution to differentiate rapid events. These values were subsequently converted to seconds for analysis.

**Status (Censoring):** This binary variable was used for survival analysis. Events were coded as uncensored (1) if the dog approached a bowl within the allotted trial time, and censored (0) if the dog failed to approach within that time.

**Demeanor:** The demeanor of the dogs was the behavioural state that they assumed during the familiarization phase. We focused exclusively on the dog's initial demeanor, displayed during the familiarization phase. This was presumed to be the innate response to unfamiliar people, prior to any habituation to the experimenters that might occur in later trials. It was classified based on the criteria described by Bhattacharjee et al. (2020). Table 1 shows descriptions of the different demeanors that were displayed by the dogs tested.

**Table 1:** Description of the different demeanors or behavioural states displayed by dogs (Bhattacharjee et al., 2020).

| Demeanor | Description |
| --- | --- |

| Affiliative | Proximity-seeking, fast or rapid tail wagging with the tail perpendicular to or below the body plane, ears pointed upward, maintaining eye contact with E1. |
|---|---|
| Anxious | Ducking posture with tail between hind legs, excessive panting, lip-licking, corners of the mouth retracted down and back. |
| Neutral | Resting without gazing at E1, lying down, or general disinterest. Approaching E1 without displaying affiliative or anxious responses were also considered within the neutral behavioural state. |

**Statistical Analysis**

All statistical analyses were conducted in R version 4.2.3. To examine the factors influencing success during the familiarization phase and the probability of making a correct choice during the test phase, we fitted generalized linear models (GLMs). For the success probability during familiarization, we tested the effects of dog sex and demeanor. We also assessed whether experimental condition influenced the likelihood of success in the test phase. To model the probability of making a correct choice, we included experimental condition, dog sex, demeanor, and approach latency as predictors. A chi-square goodness-of-fit test was used to assess differences in the dogs' bowl choices, and Cohen's *w* was calculated to estimate the corresponding effect size. To analyse factors affecting approach latency during both the familiarization and test phases, we used survival analysis. A Cox proportional hazards model was initially fitted; however, the proportional hazards assumption was violated ($p < 0.05$). Therefore, we employed a log-logistic accelerated failure time (AFT) model as an appropriate alternative for modelling approach latency. For the familiarization phase, we tested the effects of dog sex and demeanor on latency to approach. For the test phase (latency to the first and second bowl), we examined the effects of experimental condition, dog sex, and demeanor. Estimated marginal means were computed, and pairwise comparisons with Tukey's adjustment were used to identify significant differences among predictor levels for the GLMs. Statistical significance was set at α = 0.05. Model assumptions for the GLMs were evaluated using the DHARMa package. Diagnostic checks, including dispersion, zero-inflation, and residual-fitted value patterns, indicated no violations of assumptions. For the AFT models, standardized residual plots were examined and confirmed an adequate fit to the data.

**Results**

Out of the 352 adult free-ranging dogs tested, 274 were successful while 78 were unsuccessful in either the familiarization or the test phase. Out of the 78 unsuccessful experiments, 35 dogs failed in the test phase after successfully passing familiarization.

**1) Factors affecting the success of dogs in an experiment**

The model assessing factors influencing success in the familiarization phase was statistically significant (AIC = 365.68). The results indicated that sex was not a significant predictor of success (b = 0.39, SE = 0.28, z = 1.40, p = 0.162). However, demeanor was a significant predictor; compared to the reference demeanor category (affiliative), the dogs displaying an anxious demeanor were associated with a significantly lower odds of success (b = -1.01, SE = 0.31, z = -3.30, p < 0.001), as were the dogs displaying neutral demeanor (b = -0.96, SE = 0.35, z = -2.76, p = 0.006; Table S1). A Tukey-adjusted pairwise comparison confirmed that the odds of success for the affiliative dogs were 2.75 times higher than for the anxious dogs (OR = 2.748, z = 2.304, *p* = 0.003) and 2.60 times higher than for the neutral dogs (OR = 2.601, z = 2.760, *p* = 0.016). No significant difference in the odds of success was found between the anxious and the neutral dogs (OR = 0.947, z = -0.154, *p* = 0.987; Figure 1; Table S2).

In the model analysing success in the test phase, the experimental condition was not a significant predictor of overall success (*p* > 0.05; Table S3).

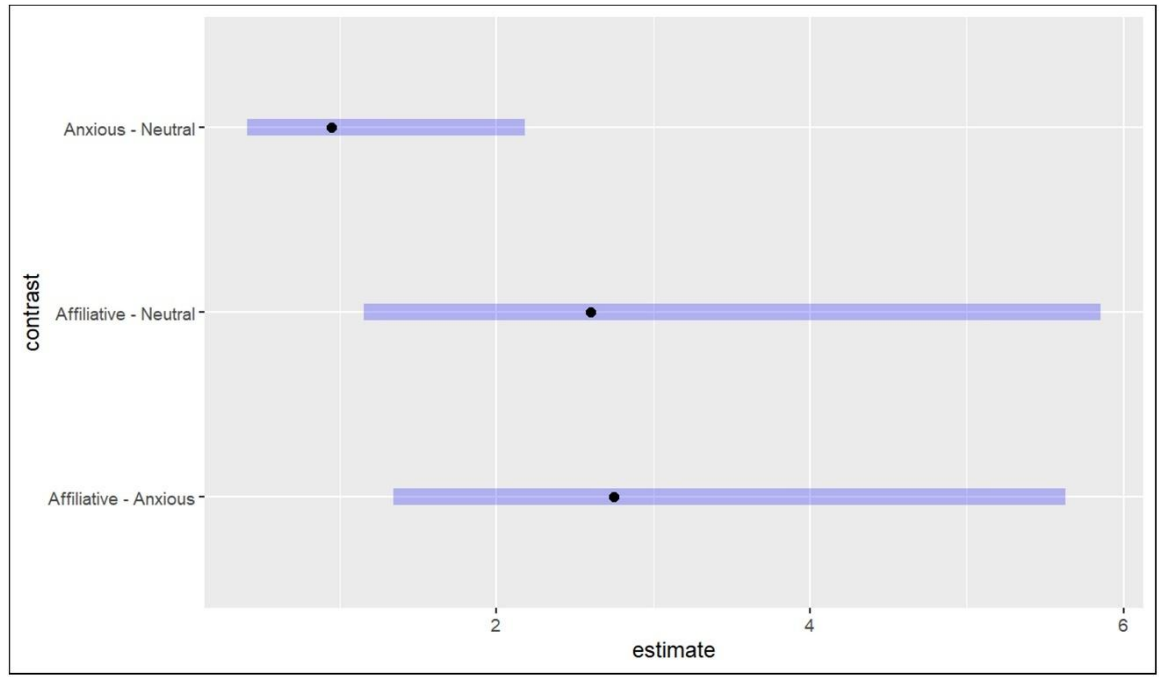

**Figure 1.** Pairwise comparisons of estimated marginal means for demeanor, with Tukey's p-value adjustment. The plot displays the odds ratios and their 95% confidence intervals for each contrast. Confidence intervals that do not cross the value 1 on the x-axis indicate a statistically significant difference between conditions (*p* < 0.05).

**2) Dog choice behaviour across six experimental conditions**

**a) Condition 1: Control**

In the control condition, there was no significant difference in the number of dogs that approached the left versus the right bowl (Chi-squared goodness-of-fit test: $\chi^2$ = 0.091, df = 1, p = 0.763; Figure 2a). The effect size, measured by Cohen's w, was 0.05 with a 95% confidence interval of 0.00 to 1.00.

**b) Condition 2: Negative control**

In the negative control condition, there was no significant difference in the number of dogs that approached the correct versus incorrect bowl (Chi-squared goodness-of-fit test: $\chi^2 = 1.200$, df = 1, p = 0.273; Figure 2b). The effect size, measured by Cohen's w, was 0.20 with a 95% confidence interval of 0.00 to 1.00.

**c) Condition 3: Pointing and gazing at the same bowl**

In the combined gazing and pointing condition, the number of dogs approaching the correct bowl was higher than chance (Chi-squared goodness-of-fit test: $\chi^2 = 11.520$, df = 1, p < 0.001; Figure 2c). The effect size, measured by Cohen's w, was 0.48 with a 95% confidence interval of 0.25 to 1.00, indicating this cue to enable more correct choices.

**d) Condition 4: Pointing-only**

In the pointing-only condition, the number of dogs approaching the correct and incorrect bowl did not differ from chance (Chi-squared goodness-of-fit test: $\chi^2 = 2.000$, df = 1, p = 0.157; Figure 2d). The effect size, measured by Cohen's w, was 0.20 with a 95% confidence interval of 0.00 to 1.00, indicating no effect of this cue in influencing correct choices.

**e) Condition 5: Gazing-only**

In the gazing-only condition, the number of dogs approaching the correct and incorrect bowl did not differ from chance (Chi-squared goodness-of-fit test: $\chi^2 = 2.880$, df = 1, p = 0.090; Figure 2e). The effect size, measured by Cohen's w, was 0.24 with a 95% confidence interval of 0.00 to 1.00, indicating no effect of this cue in influencing correct choices.

**f) Condition 6: Pointing at one bowl and gazing at the other**

In the gazing and pointing at opposite bowls condition, the number of dogs approaching the gazed versus pointed bowl did not differ from chance (Chi-squared goodness-of-fit test: $\chi^2$ = 1.280, df = 1, p = 0.258; Figure 2f). The effect size, measured by Cohen's w, was 0.16 with a 95% confidence interval of 0.00 to 1.00, indicating no effect of either of the cues in influencing correct choices.

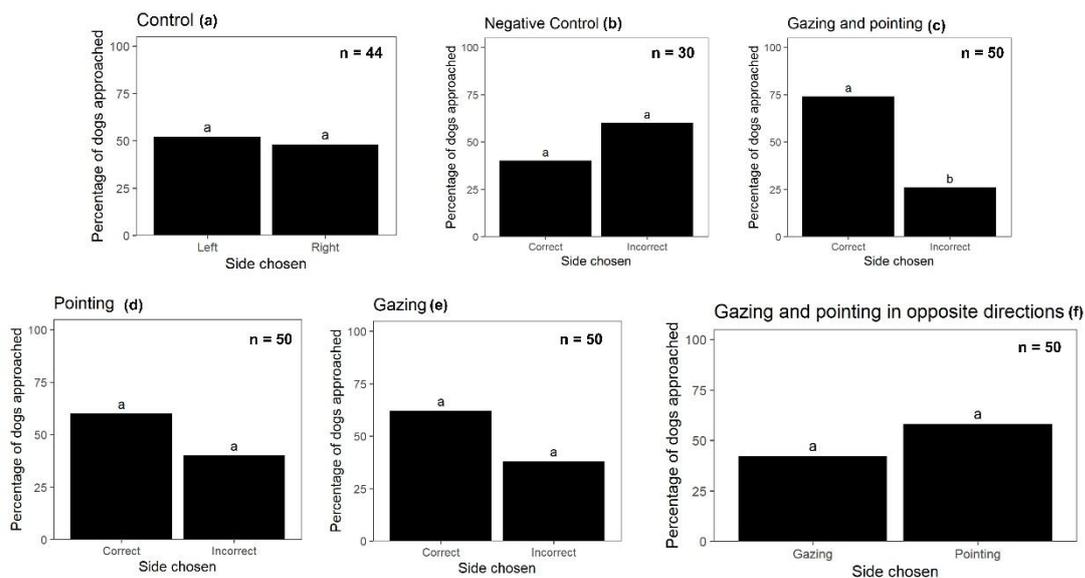

**Figure 2.** Bar graphs displaying the percentage of dogs approaching specific bowls under different test conditions. (a) Approaches to the left and right bowl during the control condition. (b) Approaches to the correct and incorrect bowl in the negative control condition. (c) Approaches to the correct and incorrect bowl during the combined gazing and pointing condition. (d) Approaches to the correct and incorrect bowl during the pointing-only condition. (e) Approaches to the correct and incorrect bowl during the gazing-only condition. (f) Approaches to the gazed versus pointed bowl when gazing and pointing were directed at opposite bowls. The number of dogs tested in each phase is denoted by "n". Letters above the bars represent statistical significance, with identical letters indicating no significant difference and different letters indicating a significant difference between groups.

## 3) Approach latency during the familiarization phase

The overall model was statistically significant ($\chi^2$ = 31.92, p < 0.001). Dog demeanor was found to be a significant predictor of approach latency during the familiarization phase. Compared to the reference demeanor group (affiliative), the anxious group had a significantly longer approach latency (AF = 2.19, z = 5.64, p < 0.001), indicating that dogs displaying an anxious demeanor took approximately 2.2 times longer to approach compared to dogs displaying affiliative demeanor. The dogs displaying a neutral demeanor also showed a significantly longer approach latency (AF = 1.39, z = 2.17, p = 0.030) which was approximately 1.4 times higher compared to the dogs displaying an affiliative demeanor. The effect of sex was not found to be a statistically significant predictor of approach latency (z = 0.94, p = 0.35; Table S4). Figure 3 shows the Kaplan-Meier curve for the effect of dog sex and demeanor on approach latency during the familiarization phase.

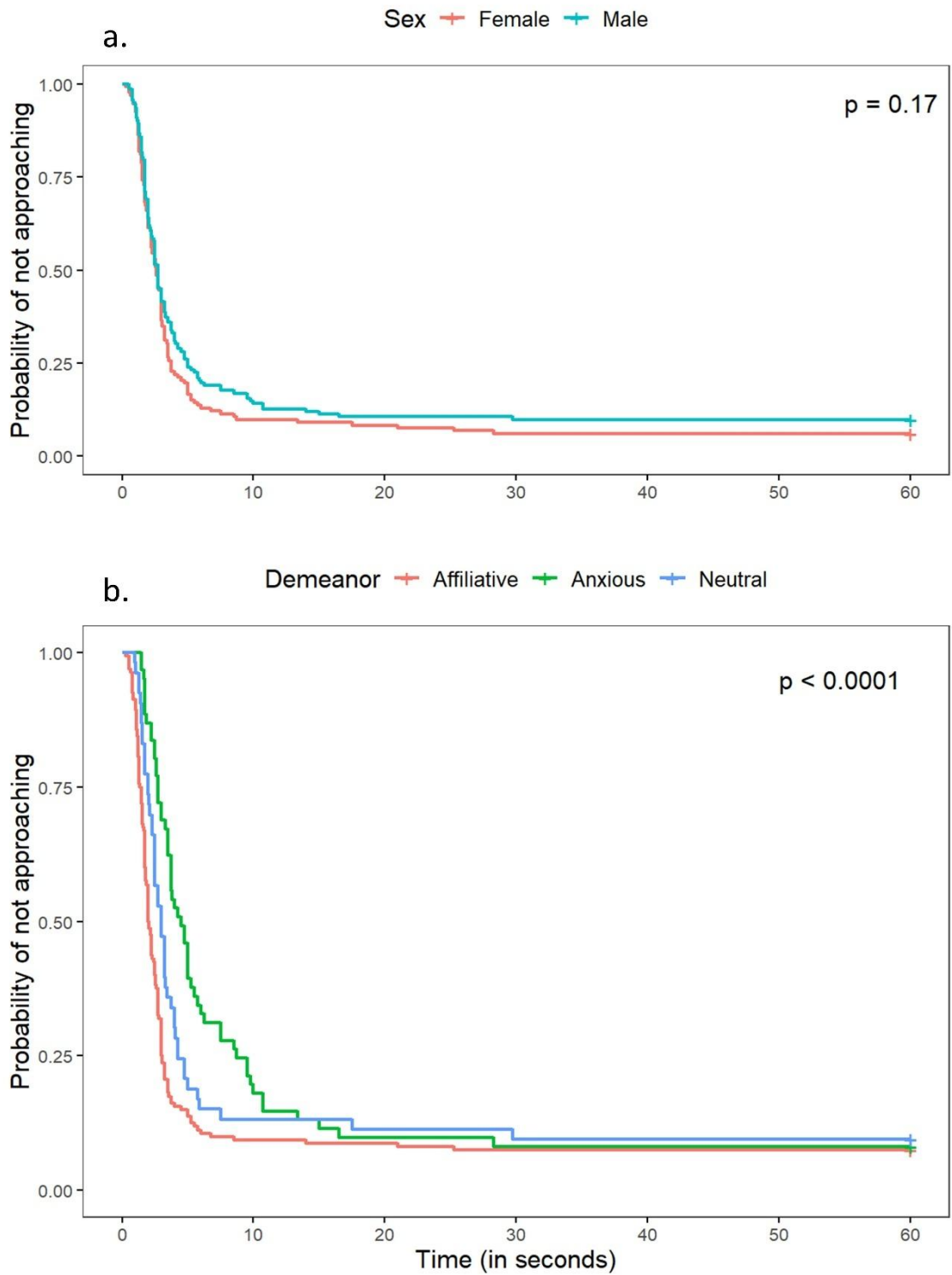

**Figure 3.** Kaplan–Meier survival curves illustrating the time to approach (seconds) during the familiarization phase. (a) The effect of sex, (b) The effect of demeanor on approach latency.

**4) Approach latency to the first bowl during the test phase**

The overall model was statistically significant ($\chi^2$ (8) = 17.630, p = 0.024). Demeanor was found to be a significant predictor of approach latency. Compared to the reference demeanor group (affiliative), the anxious dogs took significantly longer to approach the first bowl (AF = 1.82, z = 3.440, p < 0.001), which was approximately 1.8 times. Likewise, the dogs with neutral demeanor also took 1.5 times longer to approach (AF = 1.48, z = 2.130, p = 0.033). Neither sex (z = -0.410, p = 0.684) nor any of the experimental conditions (all p > 0.05; Table S5) were found to be significant predictors of approach latency to the first bowl. Figure 4 shows the Kaplan-Meier curve for the effect experimental condition, dog sex and demeanor on approach latency during the familiarization phase.

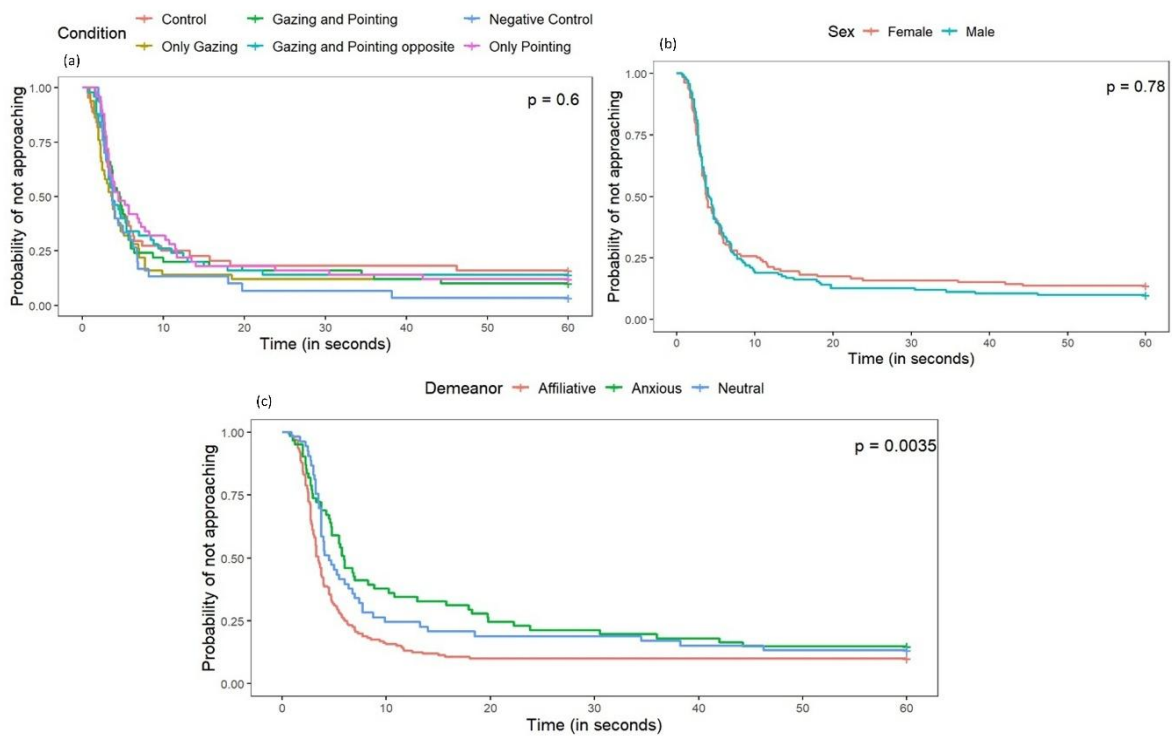

**Figure 4.** Kaplan–Meier survival curves illustrating the time to approach (seconds) to the first bowl during the test phase. (a) The effect of experimental condition, (b) The effect of sex, (c) The effect of demeanor on approach latency.

**5) Approach latency to the second bowl during the test phase**

The overall model was statistically significant ($\chi^2$ (8) = 19.320, p = 0.013). However, no individual predictors were found to be statistically significant. The approach latency of the anxious dogs approached significance (z = 1.950, p = 0.051), suggesting a potential 1.49 times acceleration in event time (AF = 1.49) compared to the reference group. No other demeanor groups (p = 0.124), sex (z = 0.260, p = 0.792), or experimental conditions (all p > 0.05; Table S6) were significant predictors of approach latency.

**6) Factors affecting the probability of making a correct choice**

The overall model was not statistically significant ($\chi^2$ (7) = 12.100, p = 0.097). Despite the non-significant overall model, the experimental condition was a key predictor. Compared to the dogs in the reference condition (negative control), the dogs in the gazing and pointing condition had 4.54 times the odds of a correct outcome (b = 1.510, SE = 0.500, p = 0.003). The dogs in the only gazing condition (OR = 2.560, b = 0.940, SE = 0.480, p = 0.051) and only pointing condition (OR = 2.320, b = 0.840, SE = 0.480, p = 0.082) did not vary significantly. Demeanor (all p > 0.05), sex (b = -0.080, p = 0.814), and approach latency (b = 0.010, p = 0.634; Table S7) were not found to be significant predictors of the outcome.

Post-hoc pairwise comparisons with Tukey's adjustment were conducted to examine differences in the odds of a correct outcome between experimental conditions. The analysis revealed that the odds of a correct outcome in the dogs receiving the gazing and pointing condition were 4.55 times higher than the odds in the dogs in the negative control condition (OR = 4.550, z = 3.024, p = 0.013). No other pairwise comparisons between conditions were found to be statistically significant (all p > 0.05; Figure 5; Table S8).

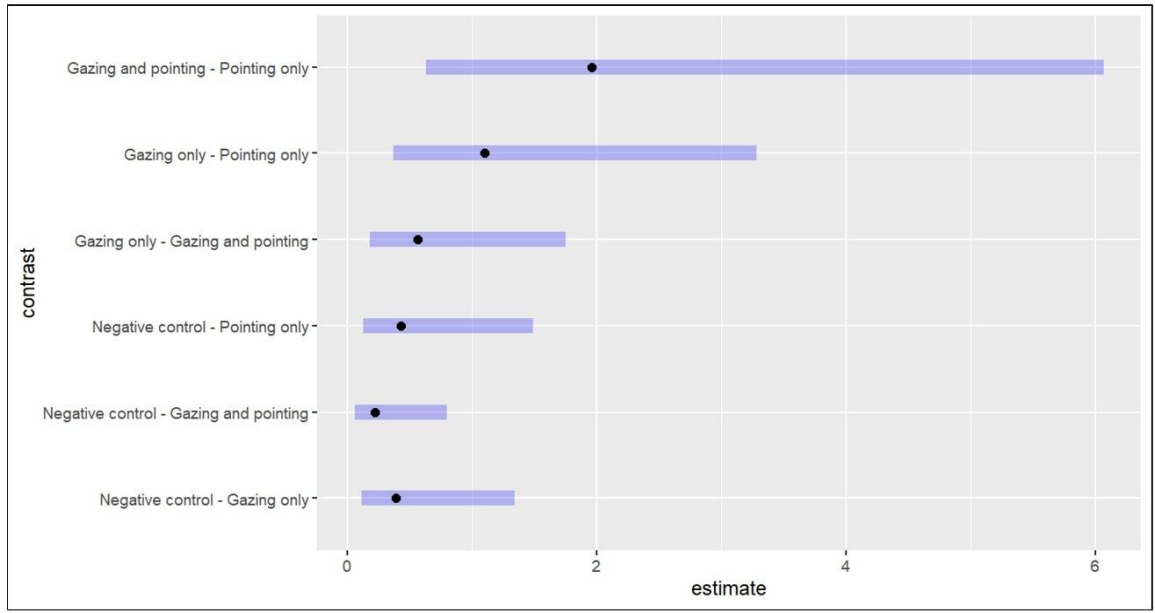

**Figure 5.** Pairwise comparisons of estimated marginal mean for experimental condition, with Tukey's p-value adjustment. The plot displays the odds ratios and their 95% confidence intervals for each contrast. Confidence intervals that do not cross the value 1 on the x-axis indicate a statistically significant difference between conditions ($p < 0.05$).

**Discussion**

This study investigated the ability of Indian free-ranging dogs to interpret human social-referential cues. Dogs classified as affiliative at the onset of the experiment significantly outperformed those that were neutral or anxious. This strongly suggests that a dog's underlying personality, particularly its social disposition towards humans, is a critical factor in its ability to succeed in tasks involving interspecies communication. We also found that for dogs that passed the familiarization phase, success in the test phase was independent of the experimental condition, suggesting the motivation of these dogs to be unaffected by the type of cue presented to them.

We then established that dogs' choices were not influenced by a pre-existing side bias (Control condition) or by olfactory cues (Negative Control condition), thereby validating our experimental setup. The primary finding revealed that dogs successfully identified the correct target only when presented with a congruent, multimodal cue combining both gazing and pointing. When cues were offered in isolation (gazing-only, pointing-only) or in conflict (gazing and pointing at opposite bowls), dogs' performance did not differ from chance. Concurrently, we found that a dog's initial demeanor towards the experimenter was a robust predictor of its approach latency, but not its choice accuracy (ability to follow human referential cues).

The success of the combined gazing and pointing condition, which resulted in a significant preference for the correct bowl with a medium-to-large effect size (Cohen's $w = 0.48$), underscores the importance of signal redundancy (Hebets et al., 2016; Akçay and Beecher, 2019) and clarity (Walsh et al., 2024) in interspecific communication. This finding aligns with a large body of literature suggesting that while dogs are highly attuned to human social cues, their comprehension is enhanced when signals are reinforcing and unambiguous (Miklósi et al., 2003; Soproni et al., 2002). The pointing gesture provides a clear directional vector, while the gazing cue adds social intent. Together, they create a powerful, easily interpretable signal that effectively guides a dog's choice. The role of gazing as an attentional cue is further supported by recent findings that gazing by humans towards free-ranging dogs while eating increases begging propensity in these dogs from unfamiliar humans (Biswas et al., 2025, preprint).

In contrast, the failure of unimodal cues (pointing-only or gazing-only) to elicit a statistically significant correct choice suggests that, for this population, isolated signals were not sufficiently salient to override chance-level performance. While dogs may perceive these individual signals, they seem to require the reinforcement of a second cue to confidently act on the information. The results from the contradictory cue condition further support this, indicating that when faced with conflicting information, these dogs do not possess a clear default strategy (e.g., "always follow the pointing").

A compelling finding is the significant effect of demeanor on approach latency. In both the familiarization phase and the initial choice during the test phase, dogs classified as anxious or neutral were significantly slower to approach than affiliative dogs. This suggests that a dog's inherent temperament is a stable predictor of its confidence or hesitation in novel situations. Anxious dogs, for instance, took approximately twice as long to approach, indicating a clear behavioural manifestation of their disposition. Interestingly, this motivation to approach was independent of the cue type, as the experimental condition did not significantly predict approach latency to the first bowl. Furthermore, sex was not found to affect approach latency or choice accuracy in any phase.

The significant effect of demeanor on approach latency disappeared when analysing the approach to the second bowl. This may be attributable to the initial interaction with the first bowl. We hypothesize that this first approach, once successfully completed, served to habituate the dogs to the experimenter and the setup. This positive or neutral experience likely mitigated the baseline hesitation in anxious and neutral individuals, resulting in subsequent approach latencies that were no longer significantly different from their affiliative counterparts.

Crucially, this documented effect of demeanor on approach latency did not translate to an effect on accuracy (ability to follow referential cue). The logistic regression model revealed that while experimental condition predicted a correct choice, demeanor did not. This decouples the dog's emotional or temperamental state from its cognitive ability to perform the task. In other words, anxious dogs were slower to make a decision, but they were no less likely to make the correct decision when the cue was clear. This highlights a critical distinction between the motivation to act and the ability to comprehend. This supports the domestication hypothesis, which posits that selection pressures during domestication conferred this baseline social-cognitive ability for cue-following across all individuals (), regardless of individual rearing; and disproves the enculturation hypothesis (), since the ability to correctly follow the clear referential cue was present even in the anxious and neutral free-ranging dogs, suggesting that this skill is a fixed, species-level trait rather than one acquired through extensive human socialization.

This study is the first to directly compare the relative effectiveness of gazing and pointing as communicative cues used by free-ranging dogs to locate hidden food, thereby extending current knowledge on referential gesture-following abilities in domestic dogs. However, it is important to note that existing literature in this area presents conflicting findings, which may be attributed to differences in the lifetime experiences of the dogs involved as well as variations in experimental methodologies. Such disparities highlight the need for careful consideration of both ontogenetic history and experimental design when interpreting results in canine communication research, and more population level studies across different socio-cultural landscapes.

In conclusion, this study reaffirms that Indian free-ranging dogs are skilled interpreters of human social communication, without any formal training, but their success is contingent on the clarity and modality of the signal. The combination of gazing and pointing appears to be a universally understood cue, while individual cues are less reliable. Furthermore, we demonstrate that a dog's demeanor is a critical factor influencing its behavioural response time, but importantly, not its cognitive ability to solve the task. These findings contribute to our understanding of the complex interplay between signal characteristics, individual temperament, and cognitive performance in the dog-human communication. Further, it highlights the need to explore the development of personalities in the free-ranging dogs, which might be shaped by their lifetime experiences with humans, and can play an important role in influencing their responses towards humans across contexts. This, in turn, has implications for the welfare of the dogs in the human-dominated habitats, and management of human-dog conflict.

**Supplementary information**

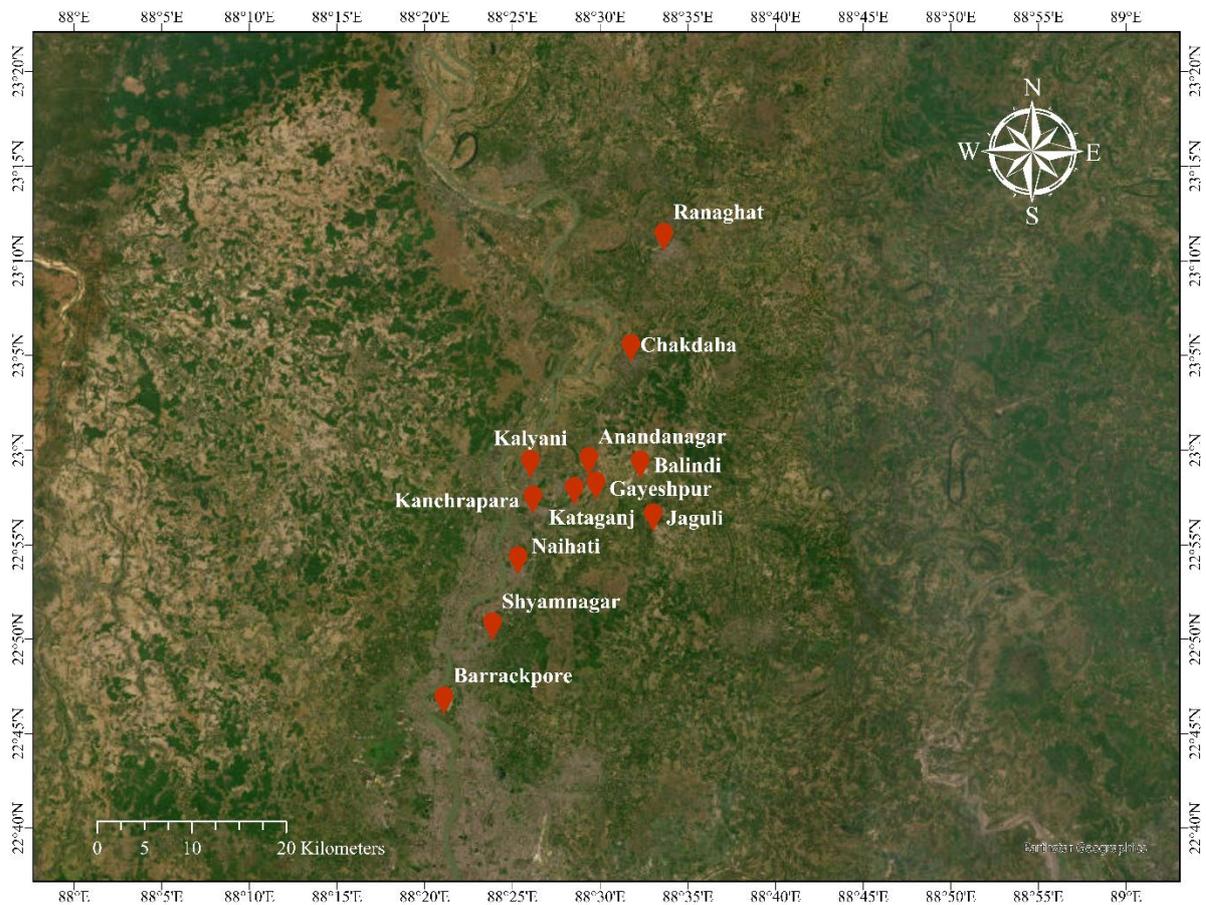

**Figure S1.** Study sites

**Table S1.** Results of the generalized linear model for the effect of sex and demeanor on success probability

glm(formula = success ~ sex + demeanor_fam, family = binomial, data = df)

|  | Estimate | Std. Error | z value | Pr(>|z|) |
|---|---|---|---|---|
| (Intercept) | 1.5891 | 0.2288 | 6.946 | 3.77e-12 *** |
| sexm | 0.3869 | 0.2766 | 1.399 | 0.161795 |
| demeanor_faman | -1.0108 | 0.3059 | -3.304 | 0.000953 *** |
| demeanor_famn | -0.9558 | 0.3463 | -2.760 | 0.005778 ** |

**Table S2.** Results of the pairwise comparisons by p value adjustment using Tukey's method for comparing the differences between demeanors

| contrast | odds.ratio | SE | df | null | z.ratio | p.value |
|---|---|---|---|---|---|---|
| af / an | 2.748 | 0.841 | Inf | 1 | 3.304 | 0.0027 |
| af / n | 2.601 | 0.901 | Inf | 1 | 2.760 | 0.0159 |
| an / n | 0.947 | 0.337 | Inf | 1 | -0.154 | 0.9869 |

**Table S3.** Results of the generalized linear model for the effect of experimental condition on success probability

glm(formula = success ~ condition, family = binomial, data = df)

|  | Estimate | Std. Error | z value | Pr(>|z|) |
|---|---|---|---|---|
| (Intercept) | 2.1748 | 0.4719 | 4.608 | 4.06e-06 *** |
| conditiong | -0.2086 | 0.6210 | -0.336 | 0.737 |
| conditiongp | 0.3510 | 0.7019 | 0.500 | 0.617 |
| conditiongp-opp | -0.3422 | 0.6064 | -0.564 | 0.573 |
| conditionnc | 1.2264 | 1.1207 | 1.094 | 0.274 |
| conditionp | -0.5623 | 0.5854 | -0.966 | 0.334 |

**Table S4.** Results of the accelerated failure time (AFT) model for the effect of sex and demeanor on approach latency during the familiarization phase

survreg(formula = Surv(latency, status) ~ sex + demeanor_fam, data = df, dist = "loglogistic")

|  | Value | Std. Error | z | p |
| --- | --- | --- | --- | --- |
| (Intercept) | 0.7457 | 0.0875 | 8.52 | < 2e-16 |
| sexm | 0.1087 | 0.1152 | 0.94 | 0.35 |
| demeanor_faman | 0.7861 | 0.1394 | 5.64 | 1.7e-08 |
| demeanor_famn | 0.3266 | 0.1504 | 2.17 | 0.03 |
| Log(scale) | -0.5908 | 0.0550 | -10.74 | < 2e-16 |

**Table S5.** Results of the accelerated failure time (AFT) model for the effect of experimental condition, sex and demeanor on approach latency to the first bowl during the test phase

survreg(formula = Surv(latency, status) ~ sex + demeanor_fam + condition, data = df, dist = "loglogistic")

|  | Value | Std. Error | z | p |
| --- | --- | --- | --- | --- |
| (Intercept) | 1.4719 | 0.1944 | 7.57 | 3.7e-14 |
| sexm | -0.0561 | 0.1377 | -0.41 | 0.68383 |
| demeanor_faman | 0.5992 | 0.1741 | 3.44 | 0.00058 |
| demeanor_famn | 0.3898 | 0.1828 | 2.13 | 0.03301 |
| conditiong | -0.2238 | 0.2309 | -0.97 | 0.33236 |
| conditiongp | 0.0247 | 0.2293 | 0.11 | 0.91416 |
| conditiongp-opp | -0.0400 | 0.2340 | -0.17 | 0.86418 |
| conditionnc | -0.0948 | 0.2531 | -0.37 | 0.70810 |
| conditionp | 0.1117 | 0.2342 | 0.48 | 0.63346 |
| Log(scale) | -0.4373 | 0.0554 | -7.89 | 3.0e-15 |

**Table S6.** Results of the accelerated failure time (AFT) model for the effect of experimental condition, sex and demeanor on approach latency to the second bowl during the test phase

survreg(formula = Surv(latency, status) ~ sex + demeanor_fam + condition, data = df, dist = "loglogistic")

|  | Value | Std. Error | z | p |
|---|---|---|---|---|
| (Intercept) | 2.3447 | 0.2269 | 10.33 | < 2e-16 |
| sexm | 0.0453 | 0.1715 | 0.26 | 0.792 |
| demeanor_faman | 0.3976 | 0.2035 | 1.95 | 0.051 |
| demeanor_famn | 0.3433 | 0.2234 | 1.54 | 0.124 |
| conditiong | -0.2203 | 0.2730 | -0.81 | 0.420 |
| conditiongp | 0.4332 | 0.2794 | 1.55 | 0.121 |
| conditiongp-opp | 0.3165 | 0.2744 | 1.15 | 0.249 |
| conditionnc | -0.5315 | 0.3331 | -1.60 | 0.111 |
| conditionp | 0.2827 | 0.2778 | 1.02 | 0.309 |
| Log(scale) | -0.2271 | 0.0548 | -4.15 | 3.4e-05 |

**Table S7.** Results of the generalized linear model for the effect of experimental condition, demeanor, sex and approach latency on probability of correct choice during the test phase

glm(formula = correct ~ condition + demeanor_fam + sex + approach_latency, family = binomial, data = df)

|  | Estimate | Std. Error | z value | Pr(>|z|) |
|---|---|---|---|---|
| (Intercept) | -0.38854 | 0.42526 | -0.914 | 0.3609 |
| conditiong | 0.93823 | 0.48024 | 1.954 | 0.0507 . |
| conditiongp | 1.51293 | 0.50032 | 3.024 | 0.0025 ** |
| conditionp | 0.84043 | 0.48293 | 1.740 | 0.0818 . |
| demeanor_faman | -0.61601 | 0.43322 | -1.422 | 0.1550 |
| demeanor_famn | 0.22158 | 0.43443 | 0.510 | 0.6100 |
| sexm | -0.07958 | 0.33829 | -0.235 | 0.8140 |
| approach_latency | 0.01014 | 0.02132 | 0.476 | 0.6342 |

**Table S8.** Results of the pairwise comparisons by p-value adjustment using Tukey's method for comparing the differences between experimental conditions

| contrast | odds.ratio | SE | df | null | z.ratio | p.value |
|---|---|---|---|---|---|---|
| nc / g | 0.391 | 0.188 | Inf | 1 | -1.954 | 0.2058 |
| nc / gp | 0.220 | 0.110 | Inf | 1 | -3.024 | 0.0133 |
| nc / p | 0.432 | 0.208 | Inf | 1 | -1.740 | 0.3027 |
| g / gp | 0.563 | 0.249 | Inf | 1 | -1.299 | 0.5635 |
| g / p | 1.103 | 0.468 | Inf | 1 | 0.230 | 0.9957 |
| gp / p | 1.959 | 0.862 | Inf | 1 | 1.529 | 0.4201 |